\documentclass[aps,pre,showpacs,floats,twocolumn,superscriptaddress,longbibliography]{revtex4-2}
\usepackage{lipsum}
\usepackage{mathrsfs}
\usepackage{bm,amsbsy,amssymb,amsmath}
\usepackage{caption}
\usepackage{subcaption}
\usepackage{graphics,graphicx,dcolumn,fleqn,epic,eepic,float,tabularx}
\usepackage{multirow,rotate,rotating,color}
\usepackage[utf8]{inputenc}
\newcommand{\figref}[1]{Fig.~\ref{#1}}
\newcommand{\eqnref}[1]{Eq.~(\ref{#1})} 
  \definecolor{tuered}{RGB}{214,0,74}
  \definecolor{tueblue}{RGB}{0,102,204}

\usepackage{tikz,pgfplots}
\usepackage{relsize}
\tikzset{fontscale/.style = {font=\relsize{#1}}}
\usetikzlibrary{calc}
\graphicspath{{img/}}

\begin{document}
\title{From suspensions to porous multilayers: microstructure formation \\ and particle packing in drying colloidal films }
  \author{Qingguang Xie}
   \email{q.xie@fz-juelich.de}
\affiliation{Helmholtz Institute Erlangen-N\"urnberg for Renewable Energy (IET-2), Forschungszentrum J\"ulich, Cauerstra{\ss}e 1, 91058 Erlangen, Germany}

\author{Jens Harting}
\email{j.harting@fz-juelich.de}
\affiliation{Helmholtz Institute Erlangen-N\"urnberg for Renewable Energy (IET-2), Forschungszentrum J\"ulich, Cauerstra{\ss}e 1, 91058 Erlangen, Germany}
\affiliation{Department of Chemical and Biological Engineering and Department of Physics, Friedrich-Alexander-Universit\"at Erlangen-N\"urnberg, Cauerstra{\ss}e 1, 91058 Erlangen, Germany}


\begin{abstract}
Drying particle suspensions is widely used to assemble particles and to fabricate porous functional layers for various applications, in which microstructural properties critically influence the overall device performance. Understanding the mechanisms governing drying-induced microstructure formation is therefore essential for predictive control of the resulting structures.
In this work, we numerically investigate the evolution of microstructures during the drying of particle suspension films, with a particular focus on the role of particle–particle interactions. For weakly interacting particles, the particles assemble into hexagonal structures at the interface, and upon drying, trigger subsequent layer-by-layer assembly.
We present a simple theoretical model to predict the time evolution of the layer thickness, validated against our simulation results. With strong particle interactions, the particles aggregate and form a network-like structure during drying, leading to a porous deposit. The porosity of the structure follows a power-law relationship with a dimensionless adhesion parameter that characterizes the particle–particle interaction force relative to the capillary force. By systematically varying the adhesion parameter, three packing regimes of the final structure are identified: hexagonal close packing, random close packing, and adhesive packing. Overall, our results demonstrate that particle–particle interactions play a decisive role in determining the final porous structure, providing practical guidance for tailoring functional layers through controlled manipulation of particle interactions.
\end{abstract}
\maketitle

\section{Introduction}

Drying of particle-based films is a common approach to assemble particles and produce functional layers for a wide range of applications, including catalyst layers in electrochemical devices such as electrolyzers~\cite{D2CS00681B,Bock2022} and fuel cells~\cite{VANDERLINDEN2023}, porous electrodes in batteries~\cite{Wei2024} and printable electronic devices~\cite{Willoughby2005, steinberger_challenges_2024}. The performance of these devices is strongly governed by the microstructural properties of the deposited layers. In particular, characteristics such as particle packing and porosity play a critical role in determining transport properties, electrochemical activity, and overall efficiency.

Despite their importance, controlling the microstructures of drying particle-based films remains a significant challenge due to the complex and coupled physical processes involved. During drying, evaporation-induced flows, interparticle interactions, and particle–substrate interactions occur simultaneously and collectively influence the final structure~\cite{Vogel2015}. These competing dynamics make it difficult to predict and precisely tailor the resulting porous morphology. Experimental studies were carried out to explore how evaporation rate, salt concentration, and the intrinsic properties of the particles (e.g., size and dispersity) shape the final deposit~\cite{HOOIVELD2025,D0SM00723D,CHEN2025}. Directly monitoring the evolution of the microstructure, however, is highly challenging, owing to the opacity of the materials and the constraints of available measurement techniques.

Consequently, numerical simulations have increasingly been used to investigate these processes, as they enable direct tracking of microstructural evolution and systematic variation of individual parameters while holding others fixed~\cite{C7SM01585B,ROUTH20042961}. For example, Jung et al.~\cite{chun_temporal_2020,Yun2025} studied the temporal evolution of microstructures in vertically drying colloidal films, focusing on the interplay among evaporation, diffusion, and sedimentation. In their model, however, colloidal interactions were represented by a simple repulsive potential, even though particle–particle interactions can play a crucial role in structure formation. More recently, Tatsumi et al.~\cite{TATSUMI2023} employed Langevin dynamics simulations to examine the influence of DLVO interactions on drying behavior. They showed that strong attractive interactions promote particle aggregation, leading to looser structures—an observation later supported by Moon et al.~\cite{moon_effect_2025} using Brownian dynamics simulations. Nevertheless, these models neglect hydrodynamic particle–particle interactions as well as capillary interactions at the interface, both of which are expected to be important in evaporating colloidal systems~\cite{Xie2026}.

In this work, we investigate the drying dynamics of thick colloidal films using a coupled lattice Boltzmann–discrete element method framework. The lattice Boltzmann method resolves fluid flow and evaporation-driven transport, while the discrete element method captures particle dynamics at the individual level. This combined approach enables direct resolution of particle–particle interactions together with particle–fluid coupling, providing a comprehensive description of microstructural evolution during drying. For weak interparticle interactions, particles self-assemble into multilayer hexagonal structures. In contrast, strong attractive interactions promote aggregation and the formation of network-like structures, resulting in highly porous deposits. To rationalize these observations, we develop a simple theoretical model that predicts the time evolution of layer thickness in the weakly interacting regime. Furthermore, by systematically varying interparticle interactions, we identify a transition between three distinct particle packing regimes: hexagonal close packing, random close packing, and cohesive packing.

\section{Methods}
\label{sec:method}
We employ the lattice Boltzmann method (LBM), a mesoscopic computational approach for modeling fluid dynamics that provides a versatile framework for simulating complex flow phenomena~\cite{bib:succi-01}. Unlike conventional computational fluid dynamics methods that directly solve the Navier--Stokes equations, the LBM is based on kinetic theory and describes fluid behavior through the evolution and collision of particle distribution functions on a discrete lattice. In the limit of small Knudsen and Mach numbers, the Navier--Stokes equations are recovered~\cite{kruger2017}.
Over the past two decades, the LBM has become a robust and widely used tool for numerical simulations of fluid flows~\cite{bib:succi-01}. The method has been extended to model multiphase and multicomponent fluids~\cite{Shan1993}, as well as suspensions of particles with varying shapes and wettabilities~\cite{ladd-verberg2001, XH21}. Its inherent parallelizability and flexibility in handling complex geometries make the LBM particularly suitable for investigating intricate fluid dynamics problems. In the following, we summarize the aspects of the method relevant to the present work and refer the reader to the literature for detailed descriptions of the methodology and implementation~\cite{Frijters2012, HXH17, Xie2026}.

We employ the pseudopotential multicomponent lattice Boltzmann method (LBM) introduced by Shan and Chen~\cite{Shan1993} with a D3Q19 lattice~\cite{Qian1992}. In this framework, two fluid components are modeled through the evolution of their respective distribution functions, which are discretized in space and time according to the lattice Boltzmann equation:
\begin{eqnarray}
  \label{eq:LBG}
  f_i^c(\mathbf{x} + \mathbf{e}_i \Delta t , t + \Delta t)&= &f_i^c(\mathbf{x},t) - \frac{\Delta t} {\tau^c} [  f_i^c(\mathbf{x},t) - \nonumber \\
  &&f_i^\mathrm{eq}(\rho^c(\mathbf{x},t),\mathbf{u}^c(\mathbf{x},t))]
  \mbox{\,,}
\end{eqnarray}
where $i=0,...,18$. $f_i^c(\mathbf{x},t)$ denotes the single-particle distribution
functions for fluid component $c=1$ or $2$, and $\mathbf{e}_i$ represents the discrete
velocity in the $i$th direction.  The parameter $\tau^c$ is the relaxation time for component
$c$ and determines its viscosity.  The macroscopic densities and velocities for
each component are defined as $\rho^c(\mathbf{x},t) = \rho_0
\sum_if^c_i(\mathbf{x},t)$, where $\rho_0$ is a reference density, and
$\mathbf{u}^c(\mathbf{x},t) = \sum_i  f^c_i(\mathbf{x},t)
\mathbf{e}_i/\rho^c(\mathbf{x},t)$, respectively. 
Here, $f_i^\mathrm{eq}$ is the second-order equilibrium distribution function defined as
\begin{eqnarray}
  \label{eq:eqdis}
  f_i^{\mathrm{eq}}(\rho^c,\mathbf{u}^c) &=& \omega_i \rho^c \bigg[ 1 + \frac{\mathbf{e}_i \cdot \mathbf{u}^c}{c_s^2} \nonumber \\
  && - \frac{ \left( \mathbf{u}^c \cdot \mathbf{u}^c \right) }{2 c_s^2} + 
  \frac{ \left( \mathbf{e}_i \cdot \mathbf{u}^c \right)^2}{2 c_s^4}  \bigg]
  \mbox{\,,}
\end{eqnarray}
where $\omega_i$ is a coefficient depending on the direction: $\omega_0=1/3$
for the zero velocity, $\omega_{1,\dots,6}=1/18$ for the six nearest neighbors
and $\omega_{7,\dots,18}=1/36$ for the nearest neighbors in diagonal direction.
$c_s = \frac{1}{\sqrt{3}} \frac{\Delta x}{\Delta t}$ is the speed of sound.

For convenience, we set the lattice spacing $\Delta x$, the timestep $\Delta t$, the reference density $\rho_0$, 
and the relaxation time $\tau^c$ to unity. This choice yields a kinematic viscosity of $\nu^c = \frac{1}{6}$ in lattice units.

The pseudopotential multicomponent model introduces a mean-field interaction force between fluid components $c$ and $\bar{c}$~\cite{Shan1993},
\begin{equation}
  \label{eq:sc}
\!\!\!\!\!  \mathbf{F}^c(\mathbf{x},t) = -\Psi^c(\mathbf{x},t) \sum_{\bar{c}} \sum_{i} \omega_i g_{c\bar{c}} \Psi^{\bar{c}}(\mathbf{x}+\mathbf{e}_i,t) \mathbf{e}_i,
\end{equation}
where $g_{c\bar{c}}$ denotes the coupling constant governing the interaction strength between the two fluid components, ultimately leading to phase separation.
We denote the surface tension of the interface by $\gamma$.
$\Psi^c(\mathbf{x},t)$ represents an ``effective mass'' and is chosen as the functional form
\begin{equation}
  \label{eq:psifunc}
  \Psi^c(\mathbf{x},t) \equiv \Psi(\rho^c(\mathbf{x},t) ) = 1 - e^{-\rho^c(\mathbf{x},t)}
   \mbox{\,.}
\end{equation}
This force $\mathbf{F}^c(\mathbf{x},t)$ is then incorporated into the dynamics of component $c$ by adding a shift
 $\Delta \mathbf{u}^c(\mathbf{x},t) =\frac{\tau^c \mathbf{F}^c(\mathbf{x},t)}{\rho^c(\mathbf{x},t)}$ 
to the velocity $\mathbf{u}^c(\mathbf{x},t)$ in the equilibrium distribution.

When the interaction parameter $g_{c\bar{c}}$ in~\eqnref{eq:sc} is chosen appropriately, the fluid components undergo phase separation, resulting in the formation of distinct phases. Each component separates into a denser majority phase with density $\rho_{ma}$ and a lighter minority phase with density $\rho_{mi}$. Owing to the diffuse nature of the interface, stress singularities at the moving contact line are naturally avoided, in contrast to sharp-interface models where such singularities commonly arise.

To induce evaporation, we impose a constant density value $\rho_H^c$ for component $c$ at the boundary sites $\mathbf{z}_H$ by prescribing the corresponding distribution functions of component $c$ as~\cite{HXH17}
\begin{equation}
f_i^c(\mathbf{z}_H,t) = f_i^\mathrm{eq}\left(\rho_H^c,\mathbf{u}^c_H(\mathbf{z}_H,t)\right).
\end{equation}
where the boundary velocity is set to $\mathbf{u}^c_H(\mathbf{z}_H,t)=0$.
When the imposed density $\rho_H^c$ is smaller than the equilibrium minority density $\rho_{mi}^c$, a density gradient develops in the vapor phase of component $c$. This gradient drives the diffusion of component $c$ toward the evaporation boundary.
The diffusion coefficient of component $c$ is given as 
 $D_c = c^{2}_{s}(\tau-\frac{1}{2}) \frac{\rho_{\bar{c}}}{\rho_c+\rho_{\bar{c}}} - \frac{c_{s}^{2}\rho_c g_{\bar{c}c}\Psi'_c\Psi_{\bar{c}}}{\rho_c+\rho_{\bar{c}}}$, 
where $\Psi'=d\Psi/d\rho$~\cite{shan_multicomponent_1995,HXH17}. It should be emphasized that our evaporation model is diffusion-dominated, as validated in our prior work~\cite{HXH17}.

The colloidal particles are discretized on the fluid lattice, and their coupling to the fluid components is enforced via a modified bounce-back boundary condition, following the approach introduced by Ladd and Aidun~\cite{ladd-verberg2001, AIDUN1998}. The particle dynamics are then described by the classical equations of motion:
\begin{eqnarray}
\mathbf{F}_{\mathrm{p}}=m \frac{d\mathbf{u}_{\mathrm{p}}}{dt} \mbox{\,} 
\label{eq: newton}
\end{eqnarray}
Here, $\mathbf{F}_{\mathrm{p}}$ denotes the total force acting on a particle with mass $m$, and $\mathbf{u}_{\mathrm{p}}$ is its velocity. The particle trajectories are updated using a leap-frog integration scheme. Since the colloids are modeled as rigid spheres, rotational dynamics and particle deformation are neglected.

The momentum exchange between particles and the surrounding fluid gives rise to hydrodynamic forces, including drag and lift contributions. The model accurately resolves lubrication interactions provided that the interparticle separation is at least one lattice spacing. For smaller separations, i.e., when the distance drops below one lattice site, a lubrication correction is introduced~\cite{ladd-verberg2001, Frijters2012}. 

In the limit of small separation $h = \left(r_{ij}-2R\right) \ll R $, 
the van der Waals force $ \mathbf{F}_{\mathrm{vdw}} $ between two spherical particles with identical radii $R$ is approximated as~\cite{HAMAKER19371058}
\begin{equation}
    \mathbf{F}_{\mathrm{vdw}} = \frac{A_H R}{12} \frac{1}{h^2} \hat{\mathbf{r}}_{ij}, \quad\mbox{for} \quad r_s \le  r_{ij}\le r_c\,,
    \label{eq:vdw}
\end{equation}
 where $A_H$ denotes the Hamaker constant and $r_c= 2R+4$ is long-range cutoff radius.  
We set the short-range cutoff radius to $r_s=2R+0.3$ in our simulations to 
eliminate the divergence of the van der Waals interaction as the particles approach contact. 

To prevent particle overlap, a Hertz potential is applied between neighboring particles~\cite{hertz1881}.
The Hertz hard-sphere potential governs particle-particle interactions at close contact, thereby removing the need for an explicit contact model.
Furthermore, to ensure that particles do not penetrate the substrate and to model particle adsorption onto the substrate, 
a Lennard-Jones (LJ) potential is introduced between particles and the substrate~\cite{Xie2026}. 

Our numerical models have been previously validated against a range of benchmark cases, including capillary interactions between neighboring particles at fluid interfaces~\cite{XDH16}, 
the evolution of the interface position during the drying of a pure liquid film and a floating droplet~\cite{HXH17}, as well as the velocity field in evaporating sessile droplets, where results were compared with theoretical predictions and experimental measurements~\cite{XH18a,schoettner2026}. We further note that the evaporation-driven dynamics in drying colloidal films are primarily controlled by vapor diffusion through the surrounding gas phase, while the detailed properties of the surrounding fluid have only a negligible effect on the overall behavior. For this reason, we employ a multicomponent formulation rather than a multiphase model, to enhance numerical stability.
 
The parameter set used in our simulations is chosen to represent colloidal particles with a characteristic radius of order $100\mathrm{nm}-1\mu m$ suspended in water. The fluid properties correspond to standard values for water, namely a dynamic viscosity of $\eta_w = 10^{-3} \mathrm{Pa\cdot s}$, a density of $\rho_w = 10^{3} \mathrm{kg/m^3}$, and a surface tension of $\sigma_w = 7.2 \times 10^{-2} \mathrm{N/m}$. We focus on a regime in which particle transport by diffusion is significantly slower than the motion of the evaporating liquid interface. This separation of time scales corresponds to a fast-evaporation limit, characterized by a large Péclet number, $Pe \gg 1$, defined as the ratio between the characteristic timescale of interface motion and that of particle diffusion. Under these conditions, Brownian motion of the colloidal particles is neglected. In addition, we assume the particle density to be comparable to that of the surrounding fluid, which renders gravitational settling of particles negligible over the timescales considered.

\section{Results}
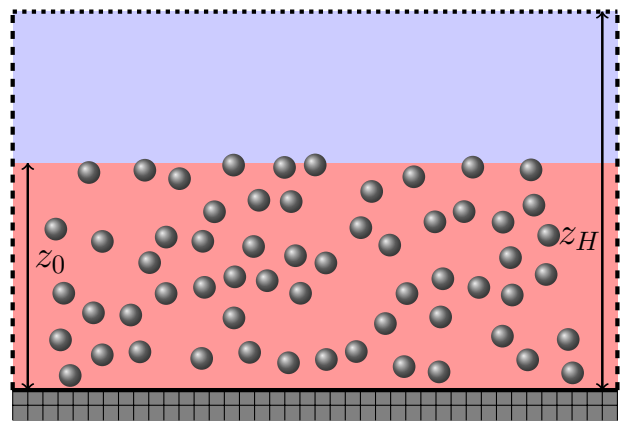
\begin{figure}[h!]
\centering
\begin{tikzpicture}

\def\radius{0.15}

\fill[red,opacity=0.4] (-4,0) rectangle (4,3);
\fill[blue,opacity=0.2] (-4,3) rectangle (4,5);

\shade[ball color=gray] (0.545559784232923,0.509514782199099) circle (\radius); 
\shade[ball color=gray] (-2.8136754076165023,1.9643179243739035) circle (\radius); 
\shade[ball color=gray] (-1.7925235851281482,2.7933830703740736) circle (\radius); 
\shade[ball color=gray] (-1.462689344788132,1.3578390192416128) circle (\radius); 
\shade[ball color=gray] (-2.1881146691867475,1.6845217662910452) circle (\radius); 
\shade[ball color=gray] (2.48712731376835,2.218678201700499) circle (\radius); 
\shade[ball color=gray] (-3.323848914448433,1.2798839600258725) circle (\radius); 
\shade[ball color=gray] (1.9719484898763193,2.3582305418313387) circle (\radius); 
\shade[ball color=gray] (0.605784494470182,2.146051673011899) circle (\radius); 
\shade[ball color=gray] (0.0020626393034848256,2.974443720851726) circle (\radius); 
\shade[ball color=gray] (-2.9911000658842806,2.874723184229879) circle (\radius); 
\shade[ball color=gray] (2.167234398955279,1.3566115344320504) circle (\radius); 
\shade[ball color=gray] (-0.8650754517075869,0.4988153384243905) circle (\radius); 
\shade[ball color=gray] (-1.9690031369603065,2.0215954091917823) circle (\radius); 
\shade[ball color=gray] (-1.4846878984090237,1.9659785716469913) circle (\radius); 
\shade[ball color=gray] (2.4790390443446757,0.7127920036540523) circle (\radius); 
\shade[ball color=gray] (-3.429193023298222,2.126931512186142) circle (\radius); 
\shade[ball color=gray] (1.2161154196074726,1.275406083888995) circle (\radius); 
\shade[ball color=gray] (2.8958417361465614,2.4434234210598875) circle (\radius); 
\shade[ball color=gray] (-0.2584039647280738,1.775517879182968) circle (\radius); 
\shade[ball color=gray] (-0.19147738663985958,1.2792492739052244) circle (\radius); 
\shade[ball color=gray] (3.0570553617965235,1.5276696657561724) circle (\radius); 
\shade[ball color=gray] (3.090776359296984,2.0444941846522804) circle (\radius); 
\shade[ball color=gray] (1.661523038810949,0.9662301921026756) circle (\radius); 
\shade[ball color=gray] (1.6908127210524029,1.471049009449061) circle (\radius); 
\shade[ball color=gray] (1.3067261493984956,2.8184149441202315) circle (\radius); 
\shade[ball color=gray] (-1.4997090521648753,0.4188457860376894) circle (\radius); 
\shade[ball color=gray] (-1.3301271557363945,2.3562133055575964) circle (\radius); 
\shade[ball color=gray] (-3.367204763743304,0.6634096059645959) circle (\radius); 
\shade[ball color=gray] (-0.8128824758158917,1.8972036882469627) circle (\radius); 
\shade[ball color=gray] (2.812830364241405,0.3998946870888721) circle (\radius); 
\shade[ball color=gray] (-2.2514273295362495,2.904815239017171) circle (\radius); 
\shade[ball color=gray] (2.608683084262756,1.2563692117416017) circle (\radius); 
\shade[ball color=gray] (2.5858365745223963,1.7505165206366549) circle (\radius); 
\shade[ball color=gray] (-1.0615601699640527,1.496378201645523) circle (\radius); 
\shade[ball color=gray] (-0.7425784474431794,2.5081073868001384) circle (\radius); 
\shade[ball color=gray] (0.7495260230613043,2.6244200437091645) circle (\radius); 
\shade[ball color=gray] (0.14963960921331054,0.40458645390503734) circle (\radius); 
\shade[ball color=gray] (1.6404705464094382,0.24086979355817167) circle (\radius); 
\shade[ball color=gray] (-1.0727199141912322,0.953692172585614) circle (\radius); 
\shade[ball color=gray] (-2.316562159013818,0.505611961256645) circle (\radius); 
\shade[ball color=gray] (-0.3507463482202593,0.3613720643112469) circle (\radius); 
\shade[ball color=gray] (-2.8149496994696763,0.4661315591282521) circle (\radius); 
\shade[ball color=gray] (2.0868471893114755,2.9452333409071785) circle (\radius); 
\shade[ball color=gray] (1.1782403260208891,0.31047336233162537) circle (\radius); 
\shade[ball color=gray] (-1.0773937261158362,2.9725894360358147) circle (\radius); 
\shade[ball color=gray] (0.9880422266549651,1.915604861253213) circle (\radius); 
\shade[ball color=gray] (-2.437693971235781,0.992035070542315) circle (\radius); 
\shade[ball color=gray] (-3.239072542390882,0.19320465454519548) circle (\radius); 
\shade[ball color=gray] (-0.4035691434142068,2.9420147337013423) circle (\radius); 
\shade[ball color=gray] (-0.3157265518093717,2.4904347585465505) circle (\radius); 
\shade[ball color=gray] (3.4075396501922715,0.22784594302901612) circle (\radius); 
\shade[ball color=gray] (-0.633079742478873,1.4455307500924697) circle (\radius); 
\shade[ball color=gray] (0.14350681903336238,1.6797751726249013) circle (\radius); 
\shade[ball color=gray] (2.8572213708543766,2.9096581919037905) circle (\radius); 
\shade[ball color=gray] (1.5862568824811323,2.223146163430898) circle (\radius); 
\shade[ball color=gray] (3.3490180149527733,0.6691746362075032) circle (\radius); 
\shade[ball color=gray] (0.9307516856286941,0.8791240842234517) circle (\radius); 
\shade[ball color=gray] (-1.9705151042590212,1.2762125793180457) circle (\radius); 
\shade[ball color=gray] (-2.931321443456701,1.0210101219032397) circle (\radius); 

\draw[ultra thick] (-4,0) -- (4,0);
\draw[ultra thick,dashed] (-4,0) -- (-4,5);
\draw[ultra thick,dashed] (4,0) -- (4,5);
\draw[ultra thick,dotted] (-4,5) -- (4,5);

\draw[thick,->] (-3.8,0) -- (-3.8,3);
\draw[thick,->] (-3.8,3) -- (-3.8,0);
\node at (-3.5,1.7) {\Large $z_0$};

\draw[thick,->] (3.8,0) -- (3.8,5);
\draw[thick,->] (3.8,5) -- (3.8,0);
\node at (3.5,2.0) {\Large $z_H$};

\fill[black,opacity=0.5] (-4,-0.4) rectangle (4,0);
\draw[step=2mm] (-4,-0.4) grid (4,0);

\end{tikzpicture}
\caption{Sketch of a thick colloidal film sitting on a substrate and covered by another fluid. The initial height of the film is $z_0$, and the distance between the evaporation boundary and the substrate is $z_H$ .}
\label{fig:film-geo}
\end{figure}
\begin{figure*}[t!]
    \centering
    \captionsetup[subfigure]{justification=centering}

 	    \begin{subfigure}{.18\textwidth}
		\includegraphics[width=0.95\textwidth]{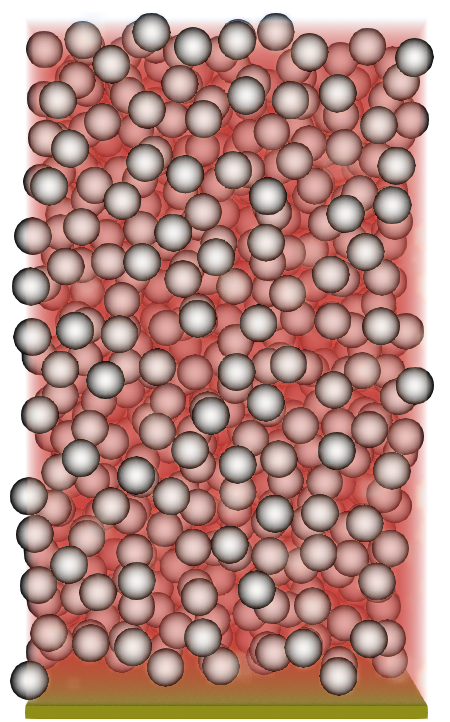}
		\subcaption{}
		\label{fig:ac_ah_low_1_side}
	\end{subfigure}
	    \begin{subfigure}{.18\textwidth}
		\includegraphics[width=0.95\textwidth]{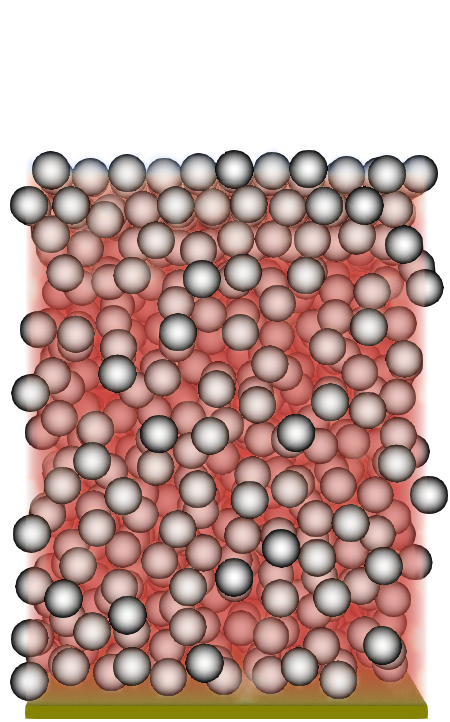}
		\subcaption{}
		\label{fig:ac_ah_low_2_side}
	\end{subfigure}
 	    \begin{subfigure}{.18\textwidth}
		\includegraphics[width=0.95\textwidth]{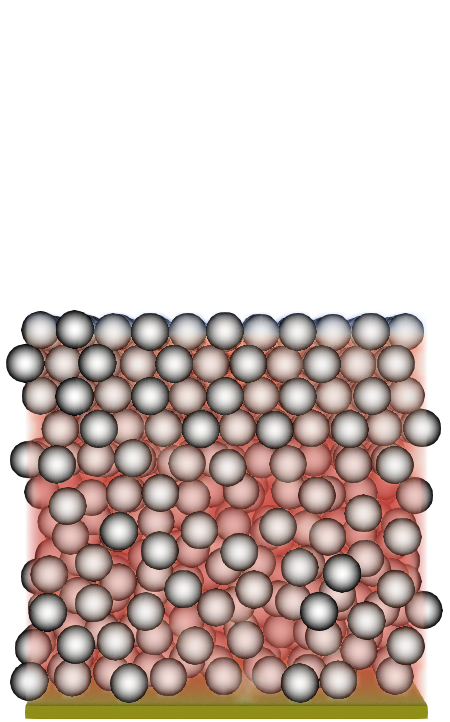}
		\subcaption{}
		\label{fig:ac_ah_low_3_side}
	\end{subfigure}
		    \begin{subfigure}{.18\textwidth}
		\includegraphics[width=0.95\textwidth]{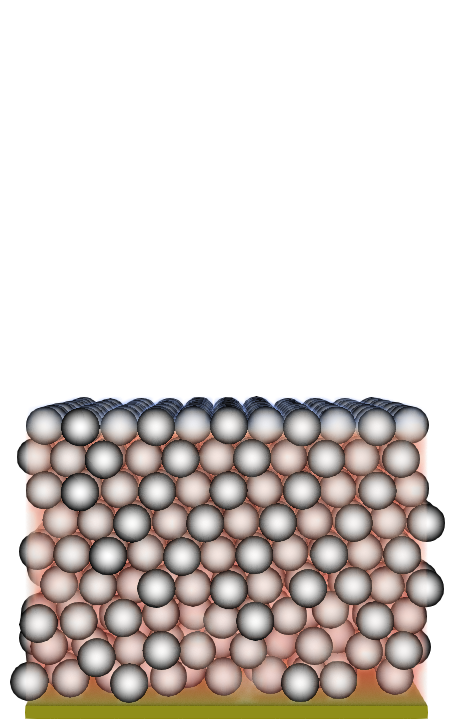}
		\subcaption{}
		\label{fig:ac_ah_low_4_side}
	\end{subfigure}
 		    \begin{subfigure}{.18\textwidth}
		\includegraphics[width=0.95\textwidth]{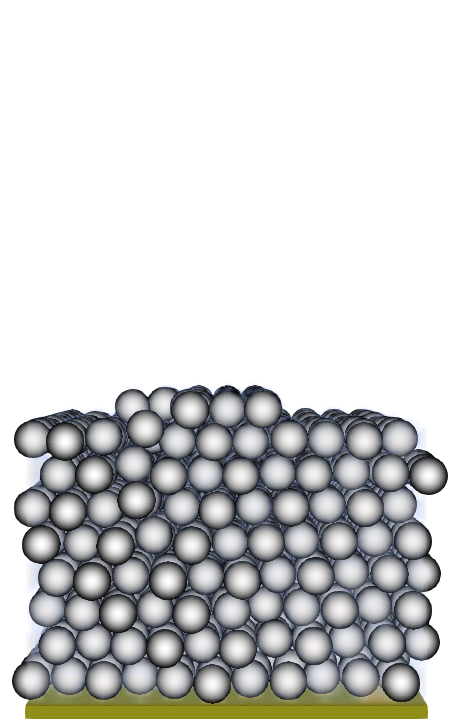}
		\subcaption{}
		\label{fig:ac_ah_low_5_side}
	\end{subfigure}
 
 	    \begin{subfigure}{.18\textwidth}
		\includegraphics[width=0.95\textwidth]{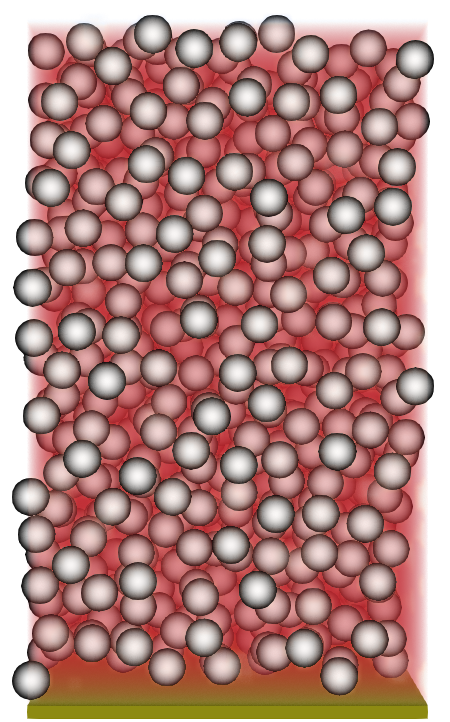}
		\subcaption{}
		\label{fig:ac_ah_low_1_side_1e-2}
	\end{subfigure}
 	    \begin{subfigure}{.18\textwidth}
		\includegraphics[width=0.95\textwidth]{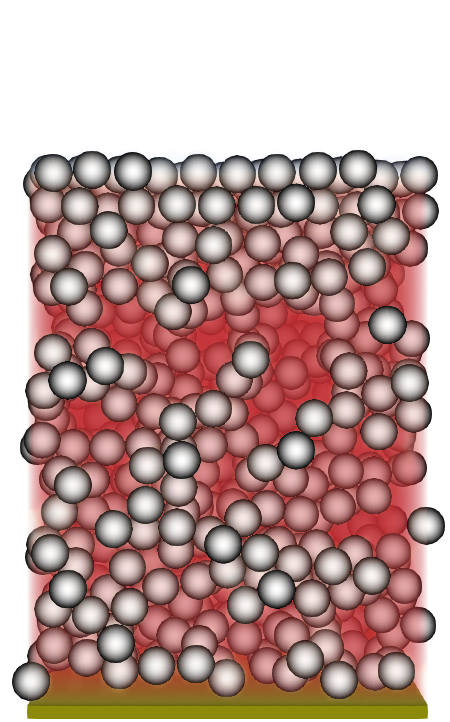}
		\subcaption{}
		\label{fig:ac_ah_low_2_side_1e-2}
	\end{subfigure}
		    \begin{subfigure}{.18\textwidth}
		\includegraphics[width=0.95\textwidth]{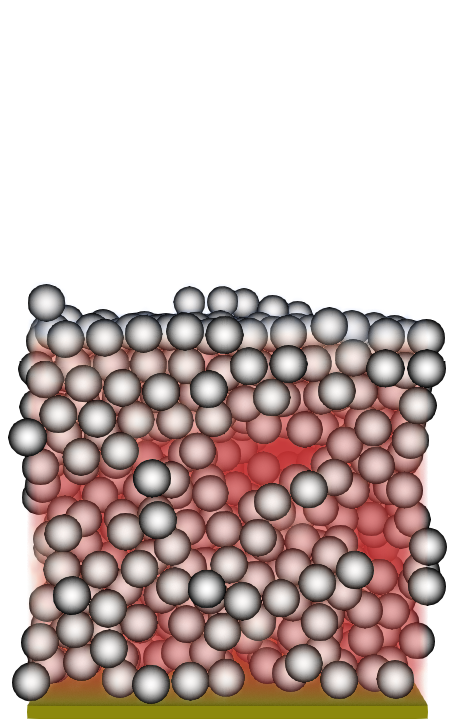}
		\subcaption{}
		\label{fig:ac_ah_low_3_side_1e-2}
	\end{subfigure}
 		    \begin{subfigure}{.18\textwidth}
		\includegraphics[width=0.95\textwidth]{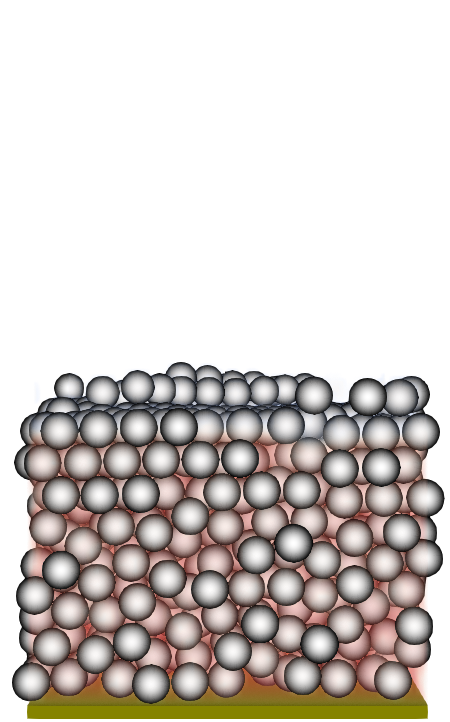}
		\subcaption{}
		\label{fig:ac_ah_low_4_side_1e-2}
	\end{subfigure}
  		    \begin{subfigure}{.18\textwidth}
		\includegraphics[width=0.95\textwidth]{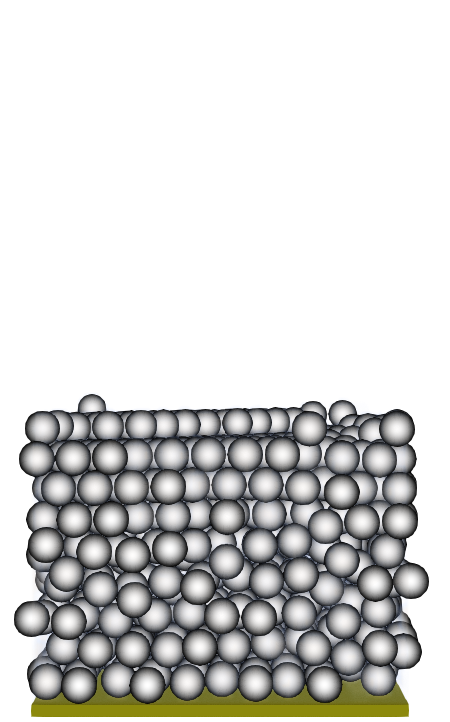}
		\subcaption{}
		\label{fig:ac_ah_low_5_side_1e-2}
	\end{subfigure}
 
 \begin{subfigure}{.18\textwidth}
		\includegraphics[width=0.95\textwidth]{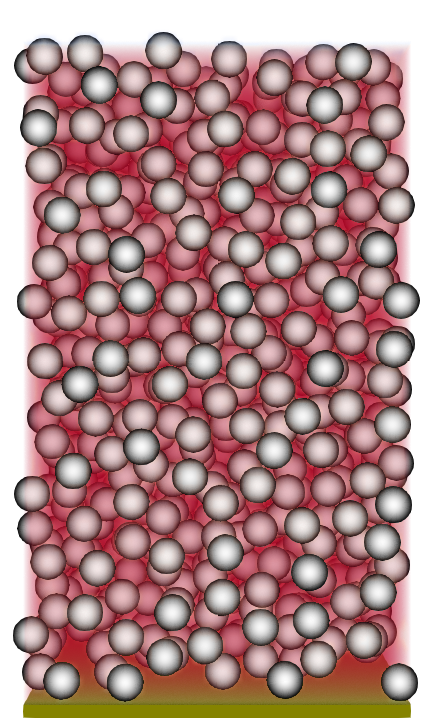}
		\subcaption{}
		\label{fig:ac_ah_low_1_side_5e-1}
	\end{subfigure}
 	    \begin{subfigure}{.18\textwidth}
		\includegraphics[width=0.95\textwidth]{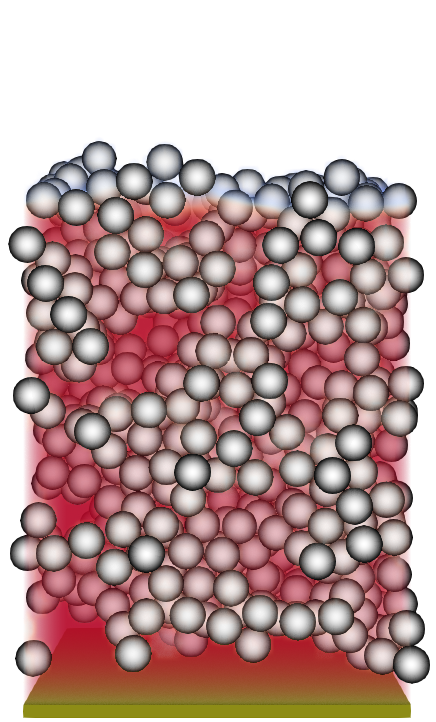}
		\subcaption{}
		\label{fig:ac_ah_low_2_side_5e-1}
	\end{subfigure}
		    \begin{subfigure}{.18\textwidth}
		\includegraphics[width=0.95\textwidth]{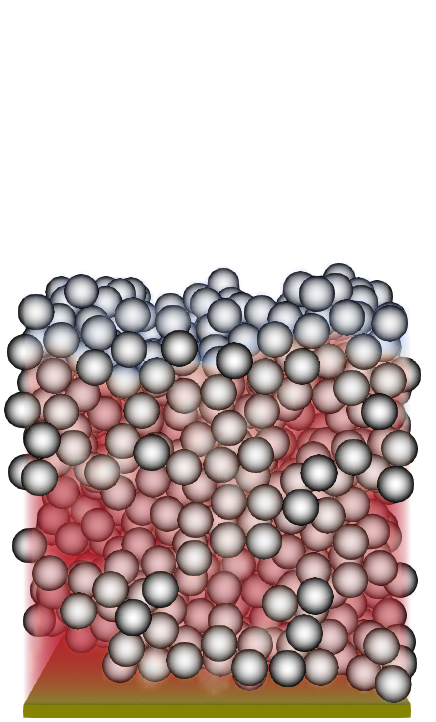}
		\subcaption{}
		\label{fig:ac_ah_low_3_side_5e-1}
	\end{subfigure}
 		    \begin{subfigure}{.18\textwidth}
		\includegraphics[width=0.95\textwidth]{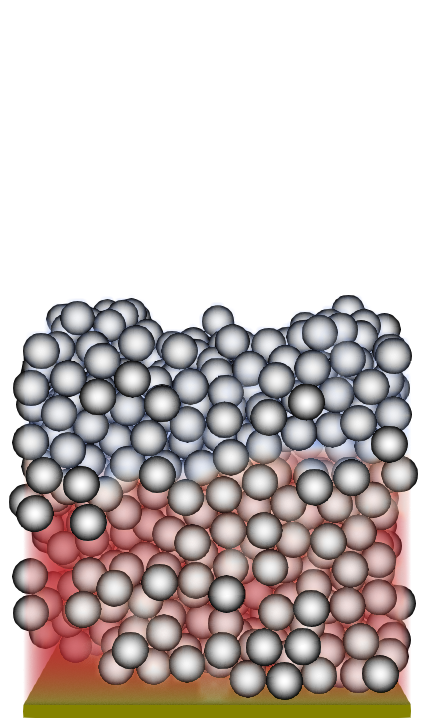}
		\subcaption{}
		\label{fig:ac_ah_low_4_side_5e-1}
	\end{subfigure}
 		    \begin{subfigure}{.18\textwidth}
		\includegraphics[width=0.95\textwidth]{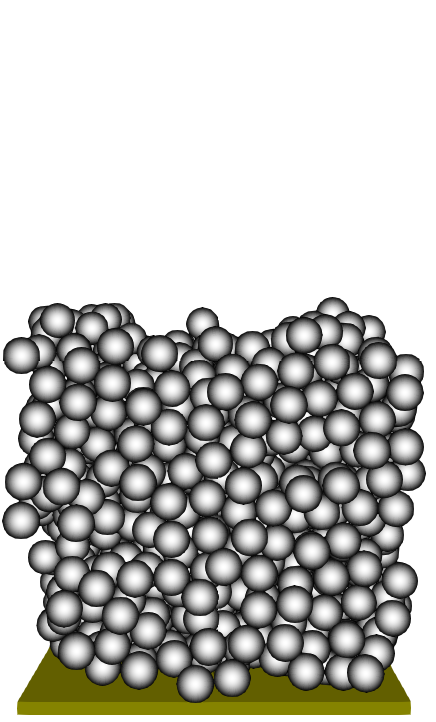}
		\subcaption{}
		\label{fig:ac_ah_low_5_side_5e-1}
	\end{subfigure}

     \caption{Snapshots of deposition processes on a substrate (light green) for different particle-particle interactions, characterized by the adhesion parameter a)-e) $\chi=0.02$ , f)-j) $\chi=0.2$  and k)-o) $\chi=10.0$. 
     The solvent is represented by the red color. 
     }
    \label{fig:snap-depo-ac}
\end{figure*}

We study the evaporation of a planar colloidal film deposited on a solid substrate, as schematically shown in~\figref{fig:film-geo}. Unless stated otherwise, all simulations are performed in a computational domain of size $64 \times 64 \times 128$ lattice nodes. Initially, the system is prepared with one region filled by fluid component $c$ and the complementary region occupied by an equally dense fluid $\bar{c}$.
This configuration gives rise to a well-defined fluid–fluid interface located at an initial position $z_0$. We identify the interface position as the position where the densities of the two components are equal, i.e., $\rho_c = \rho_{\bar{c}}$.

For the fluid–fluid interactions, we set the coupling strength in~\eqnref{eq:sc} to $g_{c\bar{c}} = 3.6$, which produces a diffuse interface spanning approximately five lattice nodes. This choice corresponds to a surface tension of $\gamma \approx 0.047$ and an effective diffusivity of $D \approx 0.12$ in lattice units.
The computational domain includes a solid substrate represented by a wall of thickness three lattice nodes located at the bottom of the system and aligned parallel to the initial interface. This wall is implemented using standard bounce-back boundary conditions. In the directions perpendicular to the substrate, periodic boundary conditions are applied. We choose particles with a radius of $3$ and include particle–particle interactions consisting of van der Waals forces, as specified in~\eqnref{eq:vdw}, and a Hertz potential, while particle–substrate interactions are modeled via a Lennard–Jones potential.

During drying, particles tend to aggregate into clusters and network-like structures due to attractive van der Waals interactions. In contrast, the descending liquid interface acts as a compressive agent: capillary forces generated by the moving interface compact the particle assemblies, often leading to partial collapse or densification of clusters and network structures. The strength of this compressive effect is primarily controlled by the surface tension and the particle size.
To quantify the competition between these two mechanisms, we introduce a dimensionless adhesion parameter $\chi = |F_{\mathrm{vdW}}^{*}| / \gamma R$, which compares the characteristic van der Waals attractive force to the capillary force scale. Here, $|F_{\mathrm{vdW}}^{*}|$ denotes the magnitude of the van der Waals force evaluated at the minimal interparticle surface separation allowed in the simulations, $h = 0.1R$, ensuring a consistent estimate of the strongest attractive interaction within the numerical framework.

The particles are randomly initialized in the solvent with a volume concentration $\phi \approx 0.2$ (\figref{fig:ac_ah_low_1_side}). Under a weak attractive interaction $\chi \approx 0.02$, the interface descends upon drying. Particles are then absorbed at the interface and reorganize into a hexagonal arrangement at the topmost layer  (\figref{fig:ac_ah_low_2_side}). This hexagonal ordering is primarily driven by confined packing of spherical particles at the interface. As drying proceeds, particles accumulate near the top interface and reorganize beneath the topmost layer. The close-packed first layer acts as a template that directs the ordered arrangement of particles in the second layer, as shown in \figref{fig:ac_ah_low_2_side} and \figref{fig:ac_ah_low_3_side}. Similarly, the subsequent layers form sequentially in a layer-by-layer manner, resulting in well-ordered multilayer structures (\figref{fig:ac_ah_low_4_side}). At the final stage, the liquid drains through the particle assembly, introducing small fluctuations and local rearrangements that lead to minor defects, as shown in \figref{fig:ac_ah_low_5_side}.

We increase the adhesion parameter to $\chi \approx 0.2$ by increasing the van der Waals interaction strength, while keeping the surface tension constant for simplicity. The same initial particle configuration is used for comparison (\figref{fig:ac_ah_low_1_side_1e-2}). Upon drying, particles are adsorbed at the interface and initially form a well-ordered layer (\figref{fig:ac_ah_low_2_side_1e-2}). However, particles in the subsurface region quickly aggregate into small clusters, as shown in \figref{fig:ac_ah_low_2_side_1e-2}. These clusters develop into supporting structures that destabilize the ordered arrangement in the upper layers (\figref{fig:ac_ah_low_3_side_1e-2} and \figref{fig:ac_ah_low_4_side_1e-2}). Meanwhile, the liquid drains through the particles, generating capillary pressure that compresses the structures and leads to densification of the clusters. Eventually, a partially ordered structure is obtained. These results indicate that stronger attractive interactions promote particle clustering beneath the interface and lead to defects in the crystallization process.

With a further increase in the adhesion parameter to $\chi \approx 10.0$, while using the same initial configuration (\figref{fig:ac_ah_low_1_side_5e-1}), particles rapidly form larger clusters at an early stage, as depicted in  \figref{fig:ac_ah_low_2_side_5e-1}. These clusters cannot be reorganized into the well-ordered arrangement favored by confined sphere packing at the interface.  
As drying proceeds, the interface descends while the liquid drains through the emerging cluster network (\figref{fig:ac_ah_low_2_side_5e-1} and \figref{fig:ac_ah_low_3_side_5e-1}). 
This cluster network is capable of withstanding the compressive effects of capillary pressure. 
Instead of forming close-packed layers, the system ultimately develops into a highly porous network-like solid structure (\figref{fig:ac_ah_low_4_side_5e-1}).

\begin{figure}[ht!]
    \centering
    \captionsetup[subfigure]{justification=centering}
    \begin{subfigure}{.45\textwidth}
		\includegraphics[width=0.99\textwidth]{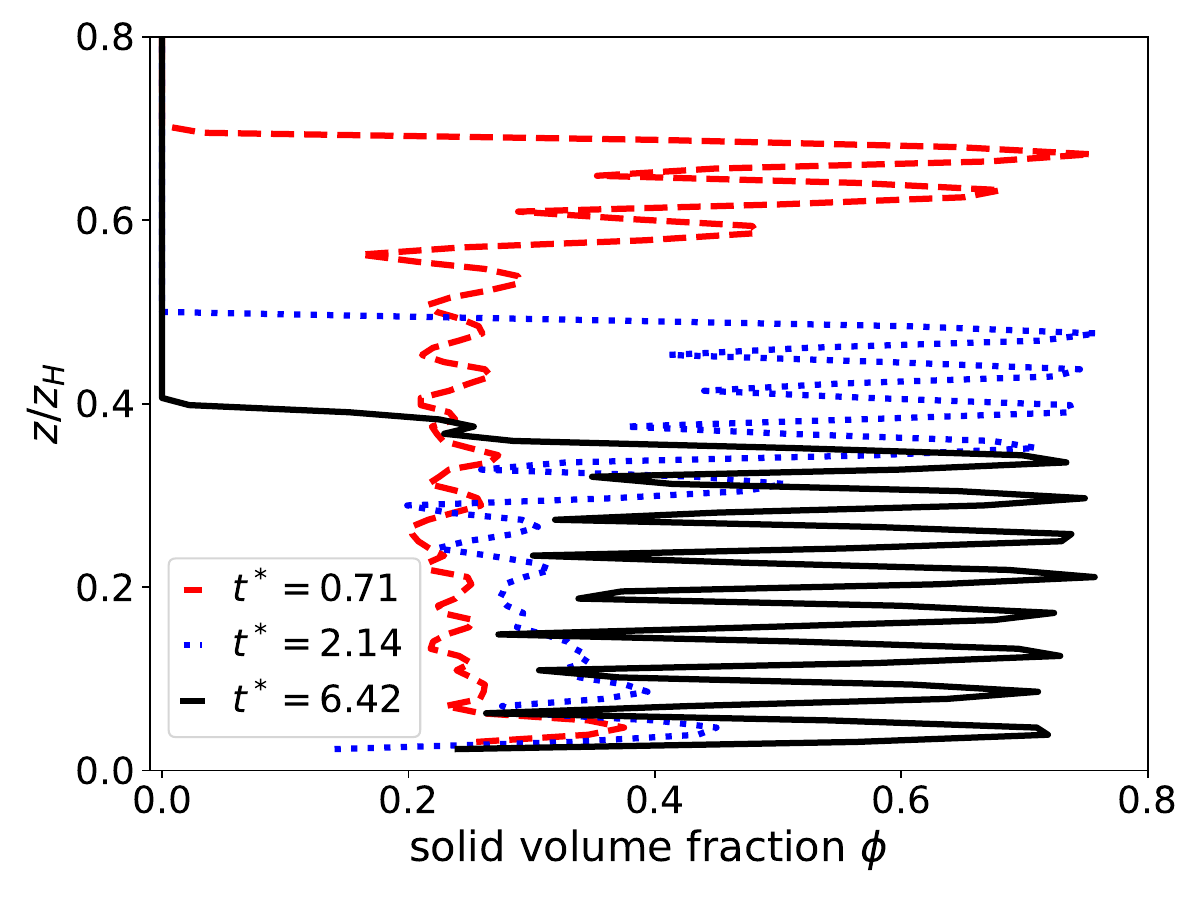}
		\subcaption{$\chi = 0.02$}
		\label{fig:vfract-z_1}
	\end{subfigure}
 
 	    \begin{subfigure}{.45\textwidth}
		\includegraphics[width=0.99\textwidth]{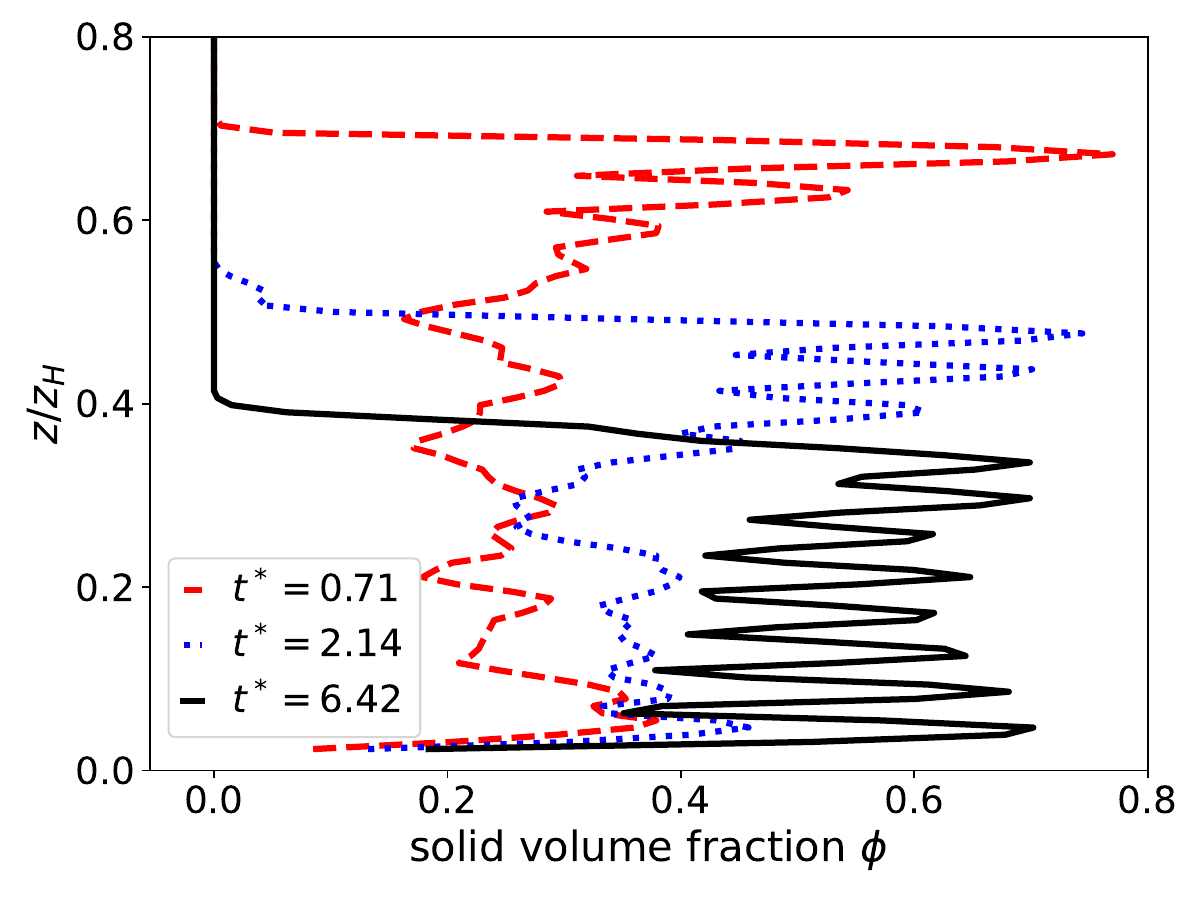}
		\subcaption{$\chi = 0.2$}
		\label{fig:vfract-z_2}
	\end{subfigure}
 
	    \begin{subfigure}{.45\textwidth}
		\includegraphics[width=0.99\textwidth]{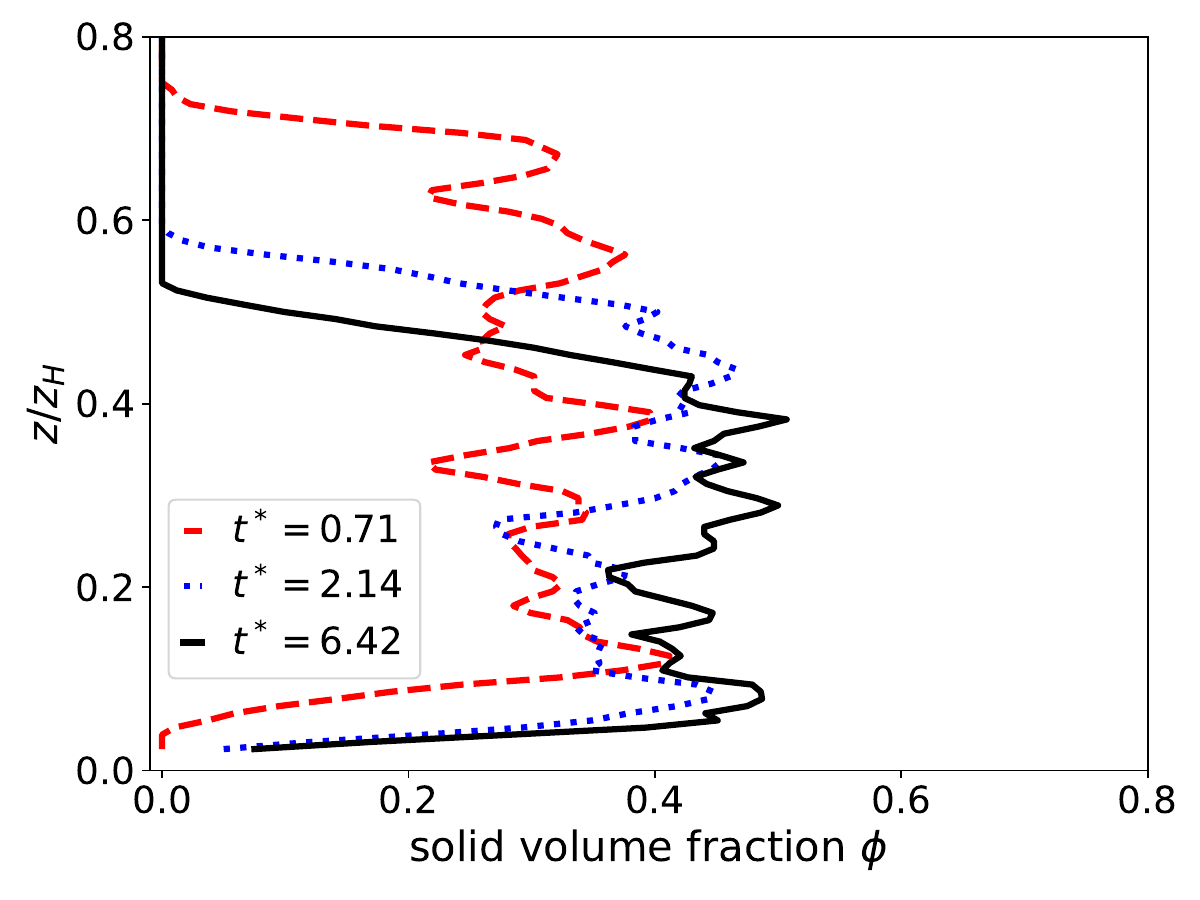}
		\subcaption{$\chi = 10.0$}
		\label{fig:vfract-z_3}
	\end{subfigure}
 
     \caption{Solid volume fraction $\phi$ along the vertical direction $z/z_H$ for different interactions a) $\chi = 0.02$,  b) $\chi = 0.2$ and c) $\chi = 10.0$. The initial particle volume fraction $\phi_0=0.2$. The normalized time $t^*$ in the legends equals $t^* = tD/L^2$.}
    \label{fig:vfract-z}
\end{figure}

The time evolution of the vertical distribution of the particle volume fraction, $\phi$, during drying at different interaction strengths is shown in~\figref{fig:vfract-z}. The particle volume fraction $\phi$ is defined as the ratio of the number of lattice sites occupied by particles to the total number of lattice sites in the $x$–$y$ plane. The normalized time $t^*$ is defined as $t^* = tD/L^2$, where $D$ is the diffusivity of the liquid and $L$ is the system size in the $x$ and $y$ directions. Under weak attractive interaction, $\chi \approx 0.02$ (\figref{fig:vfract-z_1}), at an early stage ($t^* = 0.71$), the topmost layer exhibits a peak particle volume fraction of about $0.78$, indicating a hexagonal arrangement. For reference, the ideal hexagonal packing of two-dimensional disks in a plane yields a volume fraction of approximately $0.9$. The lower value of particle volume fraction measured here is attributed to finite-size effects and the discretized representation of the particles.
Below the topmost layer, the particle volume fraction gradually (with some fluctuations) decreases toward its initial value, $\phi \approx 0.2$. This region can be identified as a transition region. Further below, the volume fraction fluctuates around the initial value, corresponding to the bulk region, where particle aggregation is absent. Near the bottom, a slight increase in particle volume fraction is observed, possibly due to confinement effects imposed by the wall. As drying proceeds, the dense region expands (e.g., $t^* = 2.14$, dotted line in \figref{fig:vfract-z_1}), indicating a layer-by-layer assembly process. At later times ($t^* = 6.42$, solid line in \figref{fig:vfract-z_1}), the system exhibits a periodic distribution of the particle volume fraction, arising from the stacking of hexagonally ordered layers.
As the adhesion parameter increases to $\chi \approx 0.2$, the particle volume fraction at the topmost layer still reaches $\phi \approx 0.78$ (dashed line in \figref{fig:vfract-z_2}). However, fluctuations in the bulk region become more pronounced, indicating the onset of particle aggregation. After drying, the particle volume fraction exhibits a periodic profile, but with larger-amplitude fluctuations (solid line in \figref{fig:vfract-z_2}), suggesting the stacking of more disordered layers. With a further increase in the adhesion parameter to $\chi \approx 10.0$, the particle volume fraction at the topmost layer decreases significantly to $\phi \approx 0.35$ (dashed line in \figref{fig:vfract-z_3}), which is much lower than the values observed at a lower adhesion parameter. In this regime, particles are unable to form a hexagonally ordered arrangement due to strong attractive interactions that promote aggregation. Interestingly, the particle volume fraction near the bottom decreases markedly. This behavior is primarily attributed to clustering and an upward migration of particles driven by interparticle attraction. While particles near the top are effectively confined at the interface, those in the lower region remain mobile due to weak particle–substrate adhesion and are therefore drawn upward into aggregates. At late times, the particle volume fraction remains highly fluctuating (dotted and solid lines in \figref{fig:vfract-z_3}), indicating the formation of a highly porous, disordered structure.

As shown in~\figref{fig:snap-depo-ac}, particles form hexagonally packed layers at the interface at a low adhesion parameter ($\chi \approx 0.02$), 
and the thickness of this region increases over time. 
To describe the temporal evolution of this hexagonally ordered region, 
we introduce a simple model based on mass conservation and liquid diffusion.
The time evolution of the interface position $z_i$ in our system can be written as~\cite{Xie2026}
\begin{equation}
    z_i = z_H - \left[(z_H-z_0)^2+2\frac{D (\rho_{mi}-\rho_{H}) }{\rho_{ma}  - \frac{\rho_{mi}-\rho_{H}}{2} } t\right]^{1/2} \mbox{.}
    \label{eq:zi}
\end{equation}
The evaporated liquid volume $V_e$ corresponding to the interface position at $z_i$ is given by
\begin{eqnarray}
   V_e &=& A( z_0-z_i) = A ( z_0-z_H) \nonumber \\ 
    && + A\left[(z_H-z_0)^2+\frac{2D (\rho_{mi}-\rho_{H}) }{\rho_{ma}  - \frac{\rho_{mi}-\rho_{H}}{2}} t\right]^{1/2} \mbox{.}
    \label{eq:evap_mass}
\end{eqnarray}
We assume that the particles corresponding to the evaporated liquid accumulate in the upper hexagonal packing region, 
and thus obtain the particle volume in this region as 
\begin{equation}
V_p =  \frac{V_e \phi_0}{(1-\phi_0)} \mbox{.}
\end{equation}
The layer thickness of the hexagonal packing region can be written as 
\begin{equation}
    L_t = \frac{V_p}{A \phi_m }  = \frac{A( z_0-z_i) \phi_0}{(1-\phi_0)A \phi_m} =  \frac{( z_0-z_i) \phi_0}{(1-\phi_0) \phi_m},
    \label{eq:layer}
\end{equation}
where $z_i$ is described by~\eqnref{eq:zi}, $\phi_0$ is the initial particle volume fraction, 
$\phi_m$ is the close packing fraction, and we take it approximately as $\phi_m=0.56$.

\begin{figure}[ht]
    \centering
    \includegraphics[width=0.45\textwidth]{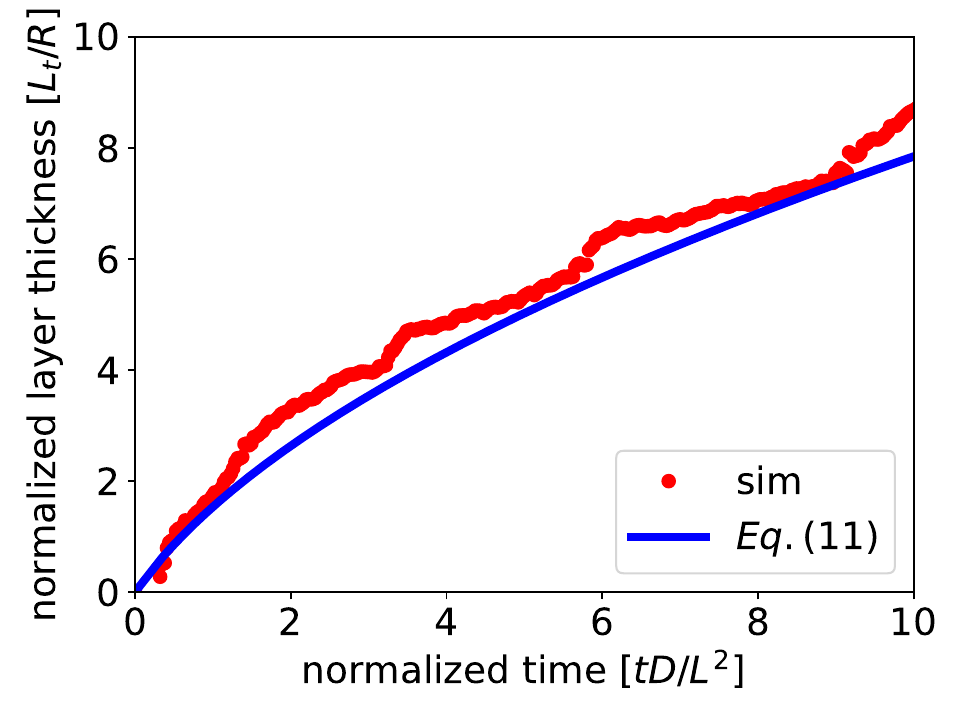}
    \caption{Time evolution of the normalized thickness $L_t/R$ of the upper hexagonal packing region with weak attractive interactions between particles ($\chi\approx 0.02$). 
    The symbols represent simulation data and the solid line corresponds to~\eqnref{eq:layer}.
}
    \label{fig:psi-chi}
    
\end{figure}

In the simulations, we define the thickness of the hexagonal packing region as the portion where the chunk-averaged volume fraction exceeds 
a threshold value of $\phi_t = 0.4$. To improve statistical accuracy, we use a chunk size of $2R$ to calculate the chunk-averaged particle 
volume fraction along the vertical direction. We first scan the chunk-averaged volume fraction profile from the top downward and define the position at which $\phi_t = 0.4$ as $z_u$. We then scan from the bottom upward and define the corresponding position as $z_d$. The layer thickness is therefore given by $L_t = z_u - z_d$. 
\figref{fig:psi-chi} shows that the analytical prediction~\eqnref{eq:layer} (solid line) agrees well with simulation results (symbols) 
for the case of weakly attractive interactions $\chi\approx 0.02$. 

Particle–particle interactions strongly influence the porosity of the resulting structures.  
In~\figref{fig:psi-chi}, we present the porosity $\psi$ as a function of the adhesion parameter, $\chi$.
For $\chi \ll 1$, the porosity varies smoothly over a relatively narrow range, as capillary interactions dominate over van der Waals interactions. In this regime, the fluid interface effectively acts as a compressive agent, promoting denser packing of particles. As $\chi$ increases into the intermediate regime ($ \chi \approx 1$), van der Waals interactions become increasingly significant and eventually dominate over capillary interactions. Correspondingly, the porosity exhibits a rapid—approximately exponential—increase with increasing~$\chi$. For $\chi \gg 1$, however, the porosity appears to approach an upper limit, suggesting a saturation behavior at a higher adhesion parameter. Motivated by this observation, we propose a power-law relationship of the form $\psi = \psi_0 + a \chi^b$ to describe the dependence of porosity on~$\chi$. The simulation data are fitted using this expression, and the resulting fit, with a power-law exponent $b = 2/3$, is shown as a solid line in~\figref{fig:psi-chi}. 

\begin{figure}[ht]
    \centering
    \includegraphics[width=0.45\textwidth]{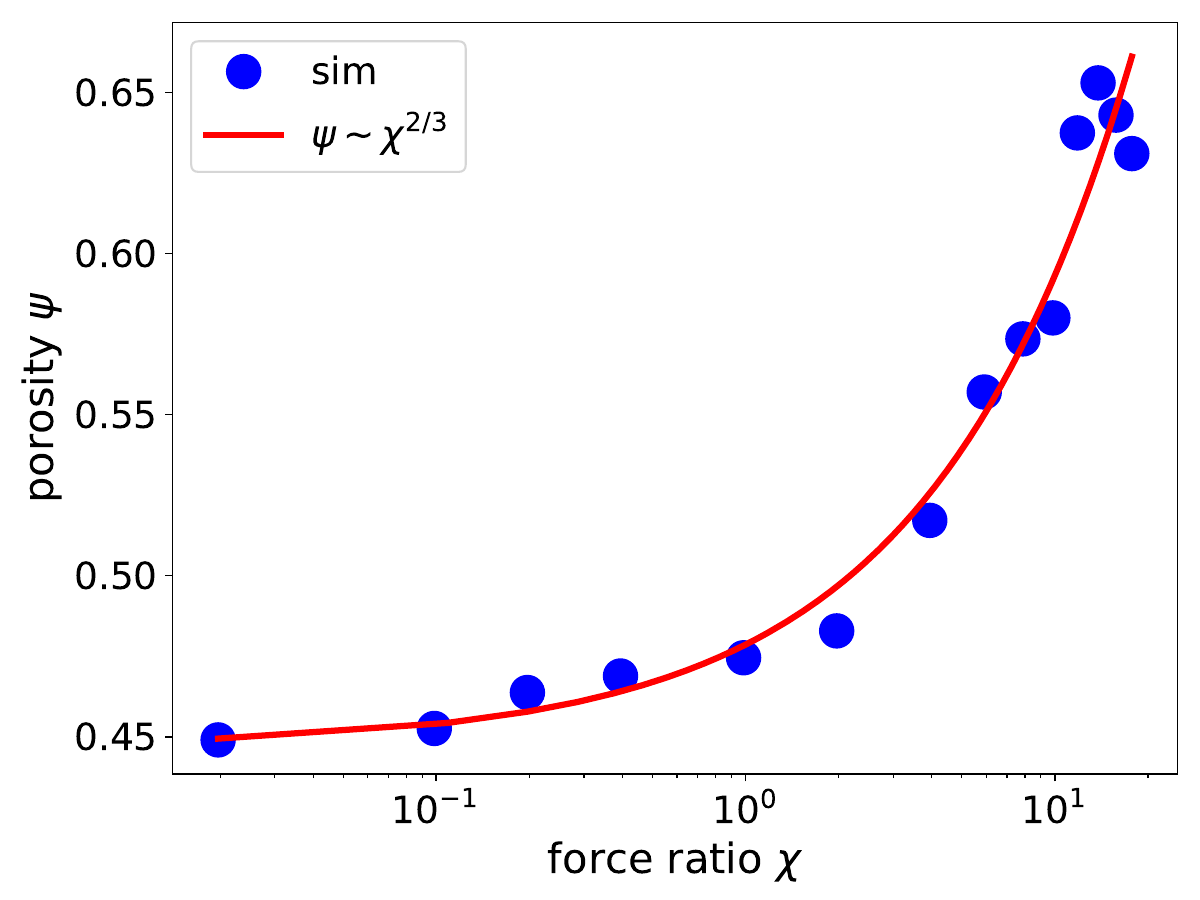}
    \caption{ Porosity $\psi$ as a function of adhesion parameter $\chi$, the interplay between van der Waals attractive force and capillary force. 
    The solid line is a power-law function fitting of simulation data (symbols). 
}
    \label{fig:psi-chi}
    
\end{figure}

\begin{figure}[t!]
    \centering
    \captionsetup[subfigure]{justification=centering}
 		    \begin{subfigure}{.32\textwidth}
   \includegraphics[width=0.97\textwidth]{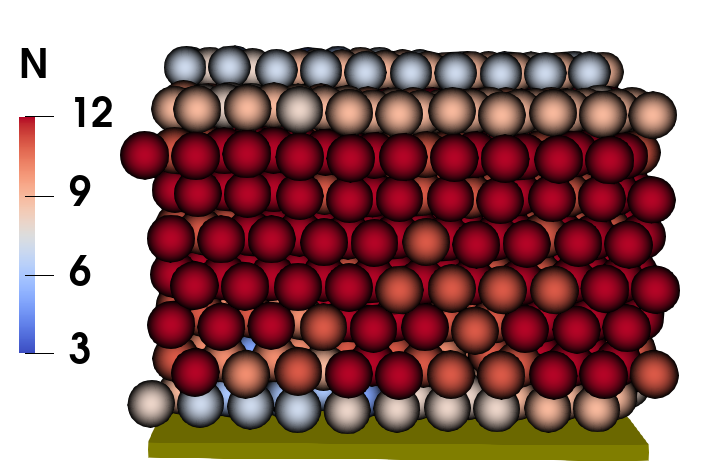}
		\subcaption{}
		\label{fig:CoNumber_1}
	\end{subfigure}
 		    \begin{subfigure}{.32\textwidth}
      \includegraphics[width=0.99\textwidth]{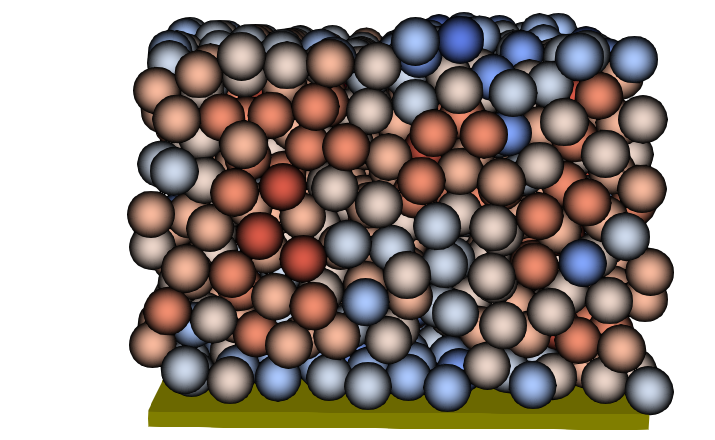}
		\subcaption{}
		\label{fig:CoNumber_2}
	\end{subfigure}
 	    \begin{subfigure}{.32\textwidth}
        \includegraphics[width=0.97\textwidth]{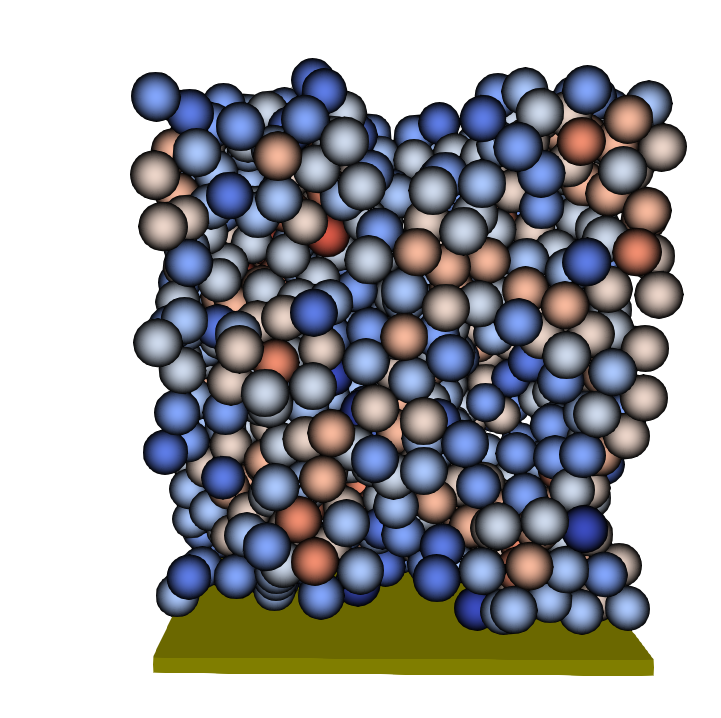}
		\subcaption{}
		\label{fig:CoNumber_4}
	\end{subfigure}

     \caption{The packing structures for different particle interaction strength: (a) hexagonal close packing with $\chi \approx 0.02$, (b) random close packing with $\chi \approx 1.0$, and (c) cohesive packing with $\chi \approx 14.0$. The color represents the coordination number $N$. }
    \label{fig:CoNumber}
\end{figure}

 The coordination number $N$ characterizes the number of contacts made by each particle. For hexagonal close packing (HCP), the coordination number is $12$: each particle is supported by three particles in the layer below, is in contact with six particles within the same layer, and supports three particles in the layer above. In contrast, for random close packing (RCP), the coordination number varies more widely, with average values of approximately $6-8$, reflecting the disordered nature of the structure~\cite{IWATA197479, YangPRE2000,Zaccone2022}. The average coordination number decreases further to a range of around $3–5$ in cohesive packing (CP)~\cite{IWATA197479, LIU2016414}. We note that the coordination number depends on the critical separation distance below which two particles are considered to be in contact. In this work, we set the critical separation distance to $R + 1$.  In \figref{fig:CoNumber}, the final packing structures for different particle interaction strengths are visualized and colored according to the local coordination number. The figure reveals a range of distinct packing regimes as the adhesion parameter increases, including HCP with minor defects (\figref{fig:CoNumber_1}, $\chi \approx 0.02$), RCP (\figref{fig:CoNumber_2}, $\chi \approx 1.0$), 
 and CP characterized by larger pores and network-like structures (\figref{fig:CoNumber_4}, $\chi \approx 14.0$). Across these regimes, the coordination number progressively decreases on average and becomes increasingly heterogeneous, reflecting the transition from ordered structures to more disordered and weakly connected networks. Consequently, these results demonstrate that a wide spectrum of packing states can be accessed by tuning the relative strength of interparticle forces, through manipulating particle-particle interaction, surface tension, or, equivalently through changes in particle size that modify the adhesion parameter.

Regarding the experimental realization of our simulation results, variations in the van der Waals interaction strength in the simulations can be mapped to the use of particles with different sizes, or different particle materials in experiments, such as silica, glass, or polystyrene. In addition, the interparticle interaction can be further tuned through changes in the surrounding medium. For example, the addition of salt or adjustments in pH can screen electrostatic repulsion or modify surface charge density, thereby effectively altering the attractive interactions. 
These mechanisms have been demonstrated experimentally to produce qualitatively different assembly behaviors and particle packings, 
ranging from dispersed structures to compact aggregates and percolated networks~\cite{LIU20221661,Mandel2024,Vogel2025}. 

\section{Conclusion}
In this work, we numerically investigate the microstructure evolution during the drying of thick colloidal particle films. 
We systematically vary the strength of particle–particle interactions and study their influence on both the dynamic evolution and the final structure. 
During drying under strong attractive interactions, particles rapidly aggregate into clusters that further connect to form an extended network. This aggregation-driven assembly results in a highly porous, loosely connected structure. Notably, the porosity of the final deposit can be described by a power-law dependence on the ratio between the van der Waals attractive force and the capillary force. In contrast, under weak attractive interactions, the drying process is dominated by front-driven ordering. Well-aligned particle layers first form at and near the interface. In this regime, particles organize into hexagonally packed layers, leading to a final more homogeneous and densely packed solid structure. We propose a simple theoretical model that captures the temporal evolution of the layer thickness, showing good agreement with the simulation results.

By systematically varying the adhesion parameter between van der Waals attraction and capillary pressure, we identify a continuous transition between three distinct packing regimes: hexagonal close packing, random close packing, and cohesive packing. This transition highlights how progressively stronger interparticle attractions drive the system from ordered crystalline-like structures toward increasingly disordered and ultimately weakly connected networks.
Overall, our results demonstrate that particle–particle interactions play a decisive role in governing drying-induced self-assembly and determine the heterogeneity and porosity of the final structure. These findings provide practical guidelines for tailoring porous materials in applications such as electrochemical energy storage, catalysis, and filtration, where transport properties are strongly linked to microstructure and porosity.

Future work will extend the present framework by incorporating additional physical effects, including sedimentation and diffusion, variations in the initial particle concentration, and particle polydispersity. Moreover, extending the model to anisotropic particles, such as ellipsoids~\cite{Arash2025}, cubes~\cite{Liu2025} and fibres~\cite{Zhao2026}, is expected to reveal a broader spectrum of structures in realistic colloidal systems.

\begin{acknowledgements}
 
We acknowledge financial support from the Deutsche Forschungsgemeinschaft (DFG, German Research Foundation) -- Project-ID 528402728 (research group ``3D-HF-MID") 
and Project-ID 416229255 (CRC 1411 ``Design of Particulate Products'').
We thank the Gauss Centre for Supercomputing e.V.(\url{www.gauss-centre.eu}) for funding this project by providing computing time
through the John von Neumann Institute for Computing (NIC) on the GCS Supercomputer JUWELS at Jülich Supercomputing Centre (JSC).
\end{acknowledgements}

\section*{Data availability}
The data generated during the current study are available at
\href{http://doi.org/10.5281/zenodo.21917418}{10.5281/zenodo.21917418}. 

\bibliographystyle{unsrt}
\bibliography{drying.bib,Ref-Xie}
\end{document}